\documentclass[a4paper,aps,prd,twocolumn,showpacs,superscriptaddress,longbibliography,nofootinbib]{revtex4-2}
\usepackage{mathrsfs,amsmath,amsthm,latexsym,amssymb,amsfonts,epsfig,cancel,enumerate,graphicx,txfonts,diagbox}
\usepackage[utf8]{inputenc}  
\usepackage[T1]{fontenc}     
\usepackage{lmodern}         
\usepackage{multirow}
\usepackage[T1]{fontenc}
\usepackage[utf8]{inputenc}
\usepackage[colorlinks=true,linkcolor=blue,citecolor=blue,urlcolor=blue]{hyperref}
\usepackage{orcidlink}

\usepackage{xcolor}
\usepackage{booktabs}
\usepackage{graphicx}
\usepackage{float}
\usepackage{subcaption}
\usepackage{placeins}

\definecolor{navy}{RGB}{0,0,150}
\usepackage{appendix}
\allowdisplaybreaks
\newcommand{\MAMO}{Department of Physics, M.A.M.O College, University of Calicut, Manassery, Calicut, 673602, Kerala, India}
\newcommand{\IMSc}{The Institute of Mathematical Sciences, C. I. T. Campus, Taramani Chennai, 600113, India}
\newcommand{\PWC}{Department of Physics, Providence Women’s College, University of Calicut, Malaparamba, Calicut, 673009, Kerala, India}
\begin{document}

\title{Strong Gravitational Lensing by Lorentzian-Euclidean Black Hole}

\author{Rukkiyya V. P\orcidlink{}}
\email{rukkiyyavp@yahoo.com}
\affiliation{\MAMO\\\PWC}

\author{Shubham Kala\orcidlink{0000-0003-2379-0204}}
\email{shubhamkala871@gmail.com}
\thanks{(Corresponding author)}
\affiliation{\IMSc}

\author{Sini R\orcidlink{}}
\email{sinir@providencecollegecalicut.ac.in}
\affiliation{\PWC}


\begin{abstract}
We investigate the strong gravitational lensing properties of the Lorentzian Euclidean black hole, a spacetime in which the horizon at $r=2M$ is not a coordinate singularity but a genuine surface of signature change, with the associated curvature singularities removed by two regularization parameters, $\rho$ and $k$. Starting from the null geodesic equations, we derive the photon sphere, the critical impact parameter, and the strong deflection limit coefficients, and use them to obtain the deflection angle and the full set of strong lensing observables, namely the angular position of the relativistic images, their angular separation, the relative magnification, and the differential time delay between successive images. We show that the photon sphere and critical impact parameter increase with $\rho$ and decrease with $k$, indicating that the two parameters have opposite effects on the optical geometry, and evaluate the resulting observables numerically for the supermassive black holes Sgr A* and M87*. Comparison with the Event Horizon Telescope shadow measurements shows that the Schwarzschild limit is mildly disfavored for Sgr A*, whereas M87* places $k$-dependent upper bounds on $\rho$, with the adopted fiducial values lying well below these limits. We further show that the shadow constrains only a combination of $\rho$ and $k$, and identify observables such as shadow circularity, higher order image time delays, and quasinormal mode spectra that are capable of breaking this degeneracy. These results identify gravitational lensing observables as effective probes of the Lorentzian--Euclidean scenario.
\end{abstract}

\maketitle

\section{Introduction}\label{sec:1}
Gravitational lensing (GL) refers to the deflection of light in the vicinity of a massive object such as a galaxy, black hole, or galaxy cluster~\cite{Turner:1984ch}. Based on the magnitude of the deflection angle, it is broadly classified into weak gravitational lensing, where the deflection is small, and strong gravitational lensing, where the deflection becomes significant~\cite{Bartelmann:1999yn}. In recent years, strong gravitational lensing has attracted considerable attention because its large deflection angles produce observable signatures that are easier to detect. As a result, it provides a powerful tool for investigating the properties of different black hole spacetimes, and extensive studies have been carried out in this area~\cite{wang2025strong,aktar2025shadows,guo2025strong,meliyeva2025theoretical,raza2026testing,molla2025observable,naskar2025strong,vachher2025strong,araujo2025gravitational,lan2025gravitational,paul2026strong,rodriguez2026strong,turakhonov2024observational,vachher2024probing,xie2024strong,junior2024gravitational,islam2024strong,Vishvakarma:2024icz}.\\
The deflection angle is a key quantity that characterizes the bending of light in a gravitational field. Although it cannot be directly measured, it plays a crucial role in determining lensing observables such as angular position, time delay, and relative magnification. Obtaining an exact expression for the deflection angle is generally challenging, and both numerical and analytical methods are employed to evaluate it. Early studies of strong gravitational lensing by compact objects with a photon sphere, including black holes and naked singularities, were carried out by Darwin \cite{darwin1959gravity}. Subsequently, Virbhadra and Ellis~\cite{virbhadra2000schwarzschild} formulated a lens equation in the strong-field limit for a Schwarzschild black hole via an asymptotically flat background metric. At the same time, Fritelli et al \cite{frittelli2000spacetime}  derived an exact lens equation expressed in integral form for Schwarzschild geometry.  A major advancement was introduced by V. Bozza \cite{bozza2002gravitational}, who developed an analytical method to calculate the deflection angle of a Schwarzschild black hole in the strong-field region, showing a logarithmic divergence in the deflection angle. This method \cite{bozza2002gravitational} has been widely applied for several spherically symmetric static metrics, such as Reissner-Nordström black holes \cite{eiroa2004strong}, Braneworld black holes \cite{whisker2005strong,eiroa2005braneworld}, and Horndeski black holes \cite{badia2017gravitational}.
Tsukamoto N \cite{tsukamoto2017deflection}  refined the strong deflection limit analysis for general asymptotically flat, static, and spherically symmetric spacetime.\\
In recent years, the strong gravitational lensing study has attracted great attention, particularly following the groundbreaking observation of the Event Horizon Telescope (EHT)~\cite{akiyama2025persistent,akiyama2019first,akiyama2022first,event2019first,event2022first,kocherlakota2021constraints,saurabh2026probing,wielgus2025millimeter}. These results have provided strong evidence for the existence of supermassive black holes at the centers of galaxies and have advanced lensing studies into the strong deflection regime beyond the weak-field approximation. They also provide a powerful framework for testing possible deviations from General Relativity by comparing theoretical predictions, such as image separation, Einstein ring radius, time delay, and shadow size, with observational constraints.\\
The Lorentzian–Euclidean black hole is a proposed solution of the vacuum Einstein field equations that seeks to resolve the problem of spacetime singularities within an extended classical framework of General Relativity \cite{capozziello2025null,bartolo2025lorentzian,de2025atemporality,battista2026shadow}. It is characterized as a complete geodesic spacetime that features a signature transition on the event horizon, which prevents causal geodesics from reaching the central singularity at r =0. The event horizon acts as a transition surface separating the Lorentzian exterior from a Euclidean-like interior region. A key feature of this geometry is atemporality, which prevents infalling matter and radiation from crossing the horizon, as their radial motion vanishes at 
r=2M and becomes non-physical beyond it. Consequently, reaching the central region would require infinite proper time, ensuring that the singularity at r=0 is never physically accessed while preserving the Schwarzschild behavior in the exterior region. Although matter and energy tend to accumulate near the horizon, the observable properties far from it remain largely unchanged, making this framework a promising candidate for describing black holes without singularities.\\
With this motivation, we investigate the strong gravitational lensing properties of a Lorentz–Euclidean black hole. The paper is organized as follows. Section~\ref{sec:2} presents the geometrical properties of the Lorentz--Euclidean black hole spacetime. Section~\ref{sec:3} presents the deflection-angle formalism. Section~\ref{sec:5} discusses the calculation of the lensing observables. Section~\ref{sec7} presents the constraints from EHT observations of Sgr A* and M87*, followed by a discussion of the observational implications and parameter degeneracy. Finally, Section~\ref{sec:6} summarizes the main results and concludes the work.

\section{Lorentzian-Euclidean black hole}\label{sec:2}
In Schwarzschild coordinates $(t,r,\theta,\phi)$, the Lorentzian--Euclidean Schwarzschild metric can be written as~\cite{Capozziello:2024ucm,DeBianchi:2025bgn,Capozziello:2025wwl}
\begin{equation}
\begin{split}
ds^{2} &= g_{\mu\nu}dx^{\mu}dx^{\nu} \\
&= -\epsilon\left(1-\frac{2M}{r}\right)dt^{2}
+ \frac{dr^{2}}{\left(1-\frac{2M}{r}\right)} \\
&\quad + r^{2}\left(d\theta^{2}+\sin^{2}\theta\, d\phi^{2}\right),
\end{split}
\label{metric1}
\end{equation}
where $M$ denotes the mass of the black hole and the function $\epsilon$ is defined as
\begin{equation}
\epsilon=\text{sign}\!\left(1-\frac{2M}{r}\right)
=2H\!\left(1-\frac{2M}{r}\right)-1,
\label{epsilon}
\end{equation}
with $H(1-2M/r)$ representing the Heaviside step function, normalized such that $H(0)=1/2$.

At the event horizon $r=2M$, the metric becomes degenerate since its determinant vanishes and the spacetime signature changes. For $r>2M$, the geometry possesses a Lorentzian signature, while for $r<2M$ the signature becomes ultrahyperbolic. In the interior region $(r<2M)$, the spacetime exhibits characteristics analogous to the Euclidean Schwarzschild geometry. This transition can be interpreted as a transformation of the temporal coordinate $t$ into an imaginary quantity when crossing the surface $r=2M$, a phenomenon that has been discussed in the context of the notion of \textit{atemporality}.

The discontinuous change of the metric signature at $r=2M$ introduces Dirac delta--like contributions in both the Ricci and Weyl components of the Riemann curvature tensor, which are formally ill-defined. To regularize these terms, one may introduce a smooth family of functions approximating the sign function. A convenient regularization is given by
\begin{equation}
\epsilon(r)=
\frac{(r-2M)^{\frac{1}{2\kappa+1}}}
{\left[(r-2M)^{2}+\rho\right]^{\frac{1}{2(2\kappa+1)}}},
\label{regularization}
\end{equation}
where $\rho$ is a small positive squared-length parameter and $\kappa$ is a positive integer satisfying $\kappa \geq 1$. In the limits of small $\rho$ and large $\kappa$, this function reproduces the sharp transition of the sign function given in Eq.~(\ref{epsilon}).
\begin{figure*}
	\centerline{
		\includegraphics[width=170mm,height=140mm]{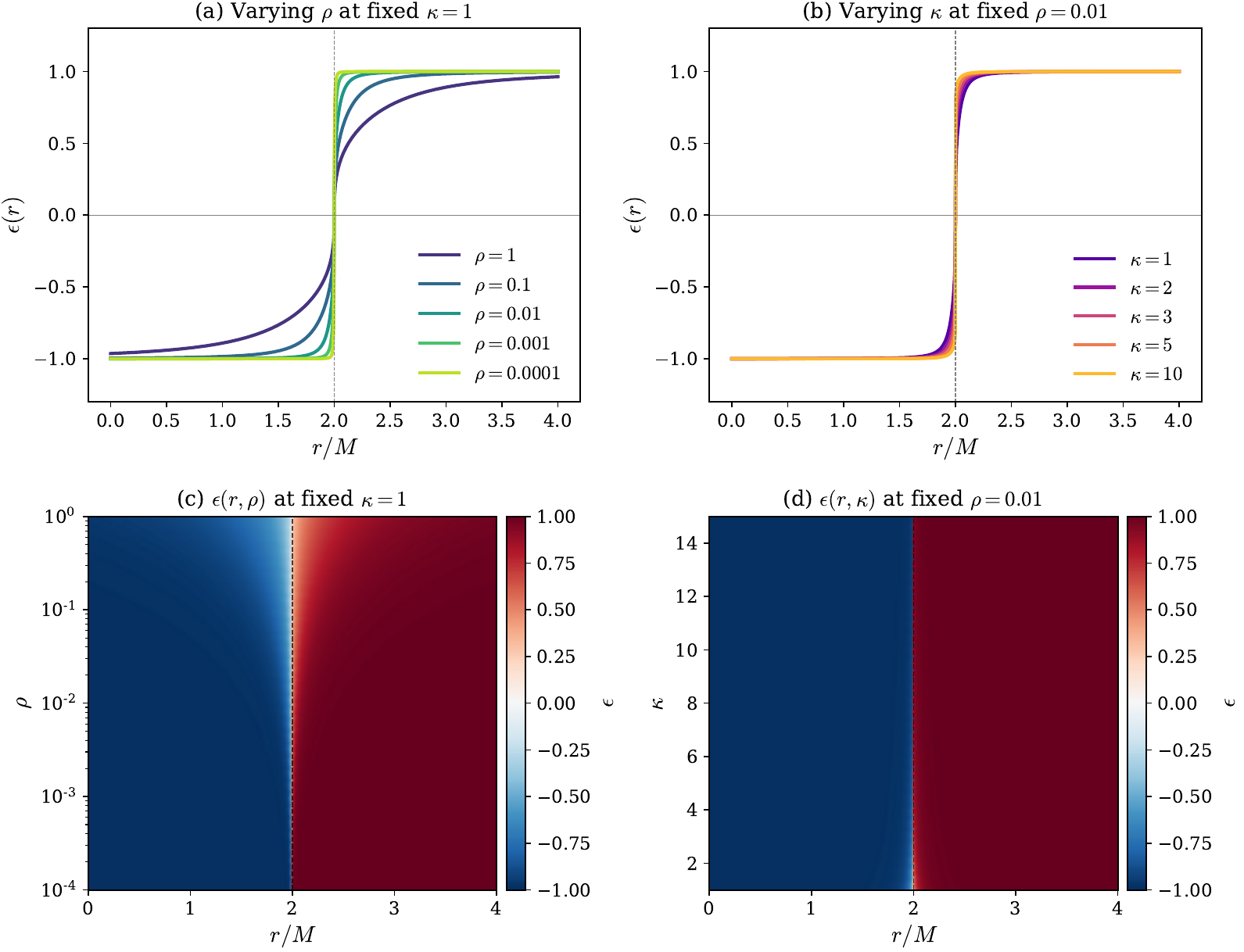}}
	\caption{Graphical interpretation of the regularization function $\epsilon(r)$ defined in Eq.~(\ref{regularization}), as a function of $r/M$. (a) $\epsilon(r)$ for fixed $\kappa=1$ and several values of $\rho$, showing that smaller $\rho$ sharpens the transition across $r=2M$ (vertical dashed line). (b) $\epsilon(r)$ for fixed $\rho=0.01$ and several values of $\kappa$, showing that larger $\kappa$ steepens the approach to $\pm1$ at fixed transition width. (c) Heatmap of $\epsilon(r,\rho)$ at fixed $\kappa=1$, with $\rho$ on a logarithmic vertical axis, illustrating the narrowing of the transition band as $\rho\to0$. (d) Heatmap of $\epsilon(r,\kappa)$ at fixed $\rho=0.01$, illustrating the same narrowing as $\kappa$ increases. In all panels, $\epsilon(r)\to\mathrm{sign}(r-2M)$ is recovered in the limits $\rho\to0$ and $\kappa\to\infty$.}
    \label{fig01}
\end{figure*}
Fig.~\ref{fig01} illustrates the behavior of the regularization function $\epsilon(r)$ defined in Eq.~(\ref{regularization}) for representative values of $\rho$ and $\kappa$. Panels (a) and (b) show that $\epsilon(r)$ smoothly interpolates between $-1$ and $+1$ across $r=2M$, with $\rho$ controlling the width of the transition region and $\kappa$ controlling the sharpness of the approach to $\pm1$ on either side; as $\rho\to0$ or $\kappa\to\infty$, $\epsilon(r)$ converges to the sharp sign function of Eq.~(\ref{epsilon}), confirming that the original Lorentzian--Euclidean metric is recovered in this limit. Panels (c) and (d) display $\epsilon(r,\rho)$ and $\epsilon(r,\kappa)$ as continuous heatmaps over the full parameter space, making explicit that the discontinuity at $r=2M$ is replaced by a finite-width band whose extent narrows monotonically as $\rho$ decreases or $\kappa$ increases.
\section{Deflection Angle of Photons in strong field limit}\label{sec:3}
Consider a line element for a spherically symmetric metric 
\begin{equation}\label{Eq.4}
    ds^2 =-A(r)dt^2+B(r)dr^2
+ C(r)(d{\theta^2}+ \sin^2{\theta}
d{\phi}^{2}).
\end{equation}
Then we derive the analytical expression for the deflection angle in the strong field limit, following the method proposed by Tsukamoto N \cite{tsukamoto2017deflection}. In this regime, there is a region where photons can approach the black hole up to a critical orbit, known as the photon sphere, which marks the closest distance a photon can reach before being captured and eventually falling into the event horizon. The radius of the photon sphere, $r_m$, is determined by the condition 
 \begin{equation}\label{Eq.26}
         \frac{C^\prime}{C}=\frac{A^\prime}{A}=0.
     \end{equation}
\begin{figure}
		\includegraphics[width=\columnwidth]{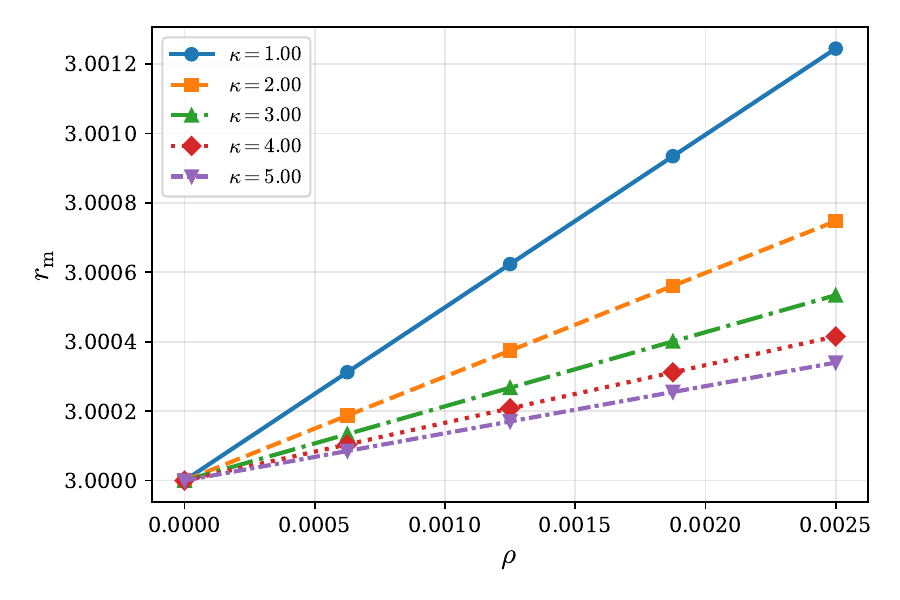}
	\caption{Variation of photon sphere as a function of the small positive squared-length parameter $\rho$ for different values of the parameter $k$.}
\label{fig1}
\end{figure}
\begin{figure}
		\includegraphics[width=\columnwidth]{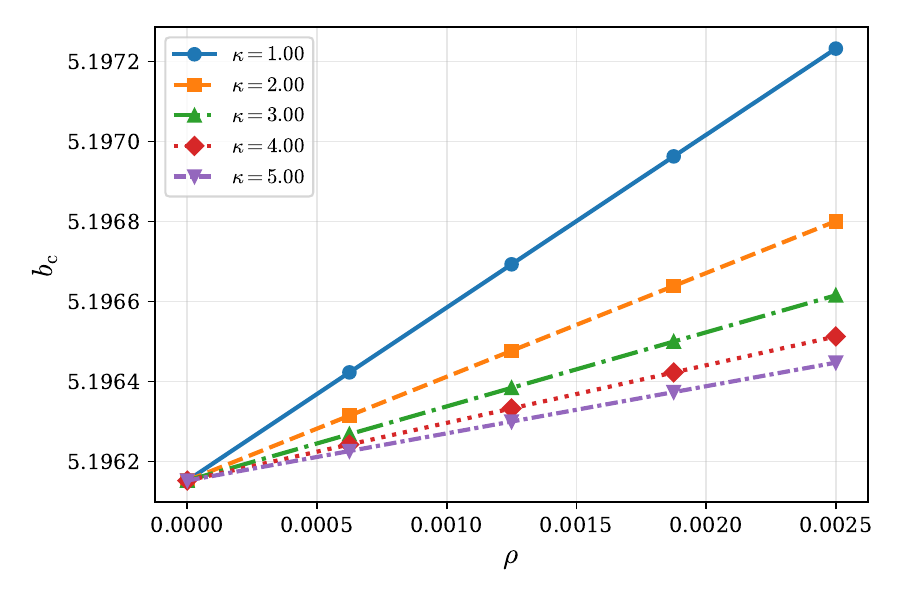}
	\caption{Variation of the critical impact parameter as a function of the small positive squared-length parameter $\rho$ for different values of the parameter $k$.}
\label{fig2}
\end{figure}
We numerically solve the above equation and plot the photon sphere radius as a function of $\rho$ for different values of $k$ (see Fig.~\ref{fig1}). The photon sphere radius is found to increase nearly linearly with increasing $\rho$, while it decreases with increasing $k$. This behavior shows that the parameter $\rho$ enlarges the region of unstable photon orbits, whereas $k$ tends to suppress it by shifting the photon sphere toward smaller radii. Since the photon sphere governs the propagation of light in the strong-gravity regime, these modifications have important consequences for black hole shadows and strong gravitational lensing observables. For $\rho=0$, the spacetime reduces to the Schwarzschild geometry, yielding the standard photon sphere radius $r_{\rm ph}=3M$.\\
The trajectory of light is described by $g_{\mu\nu} k^{\mu}k^{\nu}$ = 0 , where $k^{\mu}$ =$\dot x_{\mu}$ represents the wave number and differentiation with respect to affine parameter. Then we define two conserved quantities E and L and given by 
\begin{equation}
         E =A(r) \dot t    ,  L =C(r) \dot \phi.
\end{equation}
Now we define the impact parameter
\begin{equation}
         b=\frac{L}{E} = \frac{C(r) \dot \phi}{A(r) \dot t}. 
\end{equation}
The critical impact parameter, denoted by $b_c$, corresponds to the photon sphere radius, $r_0=r_m$, at which the photon is captured and the deflection angle diverges. It is given by
 \begin{eqnarray}\label{Eq.28}
     b_{c}=\sqrt{\frac{C_{m}}{A_{m}}}.
 \end{eqnarray}
 Fig.~\ref{fig2} shows the variation of the critical impact parameter with $\rho$ for different values of $k$. The critical impact parameter increases with $\rho$ and decreases with $k$, demonstrating that the Lorentzian--Euclidean parameters significantly alter the null geodesic structure and photon capture region of the spacetime. As a result, the black hole shadow and strong lensing observables are modified. For $\rho=0$, the Schwarzschild value $b_c=3\sqrt{3},M$ is recovered.

 The deflection angle $\alpha_{p}(r_{0})$ for a light ray arriving from infinity to the black hole is given by 
\begin{equation}
    \alpha(r_{0}) = I(r_{0}) -\pi,
\end{equation}
where $I(r_{0})$ is given by
\begin{equation}
    I(r_{0})= 2\int_{r_{0}}^{\infty} \frac{1}{\sqrt{\frac{R(r)C(r)}{B(r)}}}dr.
    \end{equation}
In order to calculate the deflection angle, we introduce a variable proposed by Tsukamoto N \cite{tsukamoto2017deflection}
\begin{equation}
    z=1-\frac{r_{0}}{r}.
\end{equation}
Therefore, $I(r_{0})$ can be written as
\begin{equation}
   I(r_{0}) = \int_{0} ^{1}f(z,r_{0})dz =\int_{0} ^{1}\frac{2r_{0}}{\sqrt{G(z,r_{0})}} dz,
\end{equation}
where the function $G(z,r_{0})$ is given by 
\begin{equation}
    G(z,r_{0})= \frac{R(z,r_{0})C(z,r_{0})}{B(z,r_{0})}(1-z)^4.
\end{equation}
The function $G(z,r_{0})$ can be expanded as 
\begin{equation}
    G(z,r_{0})=\sum _{n=1}^{\infty}c_{n}(r_{0})z^n.
\end{equation}
The expressions for $c_{1}(r_{0}) $ and $c_{2}(r_{0})$ are \cite{tsukamoto2017deflection}
\begin{equation}
    c_{1}(r_{0})=\frac{C_{0}D_{0}r_{0}}{B_{0}},
\end{equation}
and 
\begin{equation}
\begin{aligned}
c_{2}(r_{0}) &= \frac{C_{0}\,r_{0}}{B_{0}}
\Biggl\{
D_{0}\left[\left(D_{0}
- \frac{B_{0}^{\prime}}{B_{0}}\right)r_{0}-3\right] \\
&\qquad\qquad
+ \frac{r_{0}}{2}
\left(
\frac{C_{0}^{\prime\prime}}{C_{0}}
-
\frac{A_{0}^{\prime\prime}}{A_{0}}
\right)
\Biggr\}.
\end{aligned}
\end{equation}
In the  strong field limit  $r_{0} \rightarrow r_{m}$, the coefficient $c_{2}$ reduces to
\begin{equation}
c_{2}(r_{m}) =\frac{C_{m}r_{m}^2}{2B_{m}}D_{m}^\prime,
\end{equation}
where 
\begin{equation}
    D_{m}^\prime= \frac{C_{m}^{\prime\prime}}{C_{m}}-\frac{A_{m}^{\prime\prime}}{A_{m}}.
\end{equation}
To evaluate $I(r_{0})$, we decompose the integral into two parts,
\begin{equation}
    I(r_{0}) = I_{R}(r_{0}) + I_{D}(r_{0}),
\end{equation}
where $I_{D}(r_{0})$ is defined as
\begin{equation}
    I_{D}(r_{0}) = \int_{0}^{1}f_{D}(z,r_{0})dz,
\end{equation}
with
\begin{equation}\label{Eq.39}
    f_{D}(z,r_{0}) =\frac{2r_{0}}{\sqrt{c_{1}(r_{0}z+c_{2}(r_{0}z^2}}.
\end{equation}
Integrating Eq.~(\ref{Eq.39}) yields the closed-form expression
\begin{equation}
    I_{D}(r_{0}) =\frac{4r_{0}}{\sqrt{c_{2(r_{0})}}}\log{\frac{\sqrt{c_{2}(r_{0})}+\sqrt{c_{1}(r_{0})+c_{2}(r_{0})}}{\sqrt{c_{1}(r_{0})}}}.
\end{equation}
The regular part $I_{R}(r_{0})$ is given by
\begin{equation}
    I_{R}(r_{0}) = \int_{0}^{1}f_{R}(z,r_{0})dz,
\end{equation}
where
\begin{equation}
    f_{R}(r_{0}) = f(z,r_{0})-f_{D}(z,r_{0}).
\end{equation}
In the strong-field limit $b \rightarrow b_{c}$, where $b$ denotes the impact parameter, the deflection angle $\alpha(b)$ can be determined following the method of Ref.~\cite{tsukamoto2017deflection} as
\begin{equation}
    \alpha(b) = -\bar{a}\log\!\left(\frac{b}{b_{c}}-1\right) 
    + \bar{b} 
    + \mathcal{O}\!\left((b-b_{c})\log(b-b_{c})\right).
\end{equation}
The strong-deflection-limit coefficients $\bar{a}$ and $\bar{b}$ are expressed as
\begin{eqnarray}
       \overline{a}=\sqrt{\frac{2B_{m} C_{m}}{C_{m}^{\prime\prime}A_{m} - A_{m}^{\prime\prime} C_{m}}},
   \end{eqnarray}
\begin{equation}\label{Eq.45}
    \bar{b} = \bar{a}\log\!\left(r_{m}^{2}
    \left[\frac{C_{m}^{\prime\prime}}{C_{m}} 
    - \frac{A_{m}^{\prime\prime}}{A_{m}}\right]\right) 
    + I_{R}(r_{m}) - \pi,
\end{equation}
where the subscript $m$ denotes quantities evaluated at $r = r_{m}$, the radius of the photon sphere.

\begin{figure*}
	\centerline{
		\includegraphics[width=170mm,height=70mm]{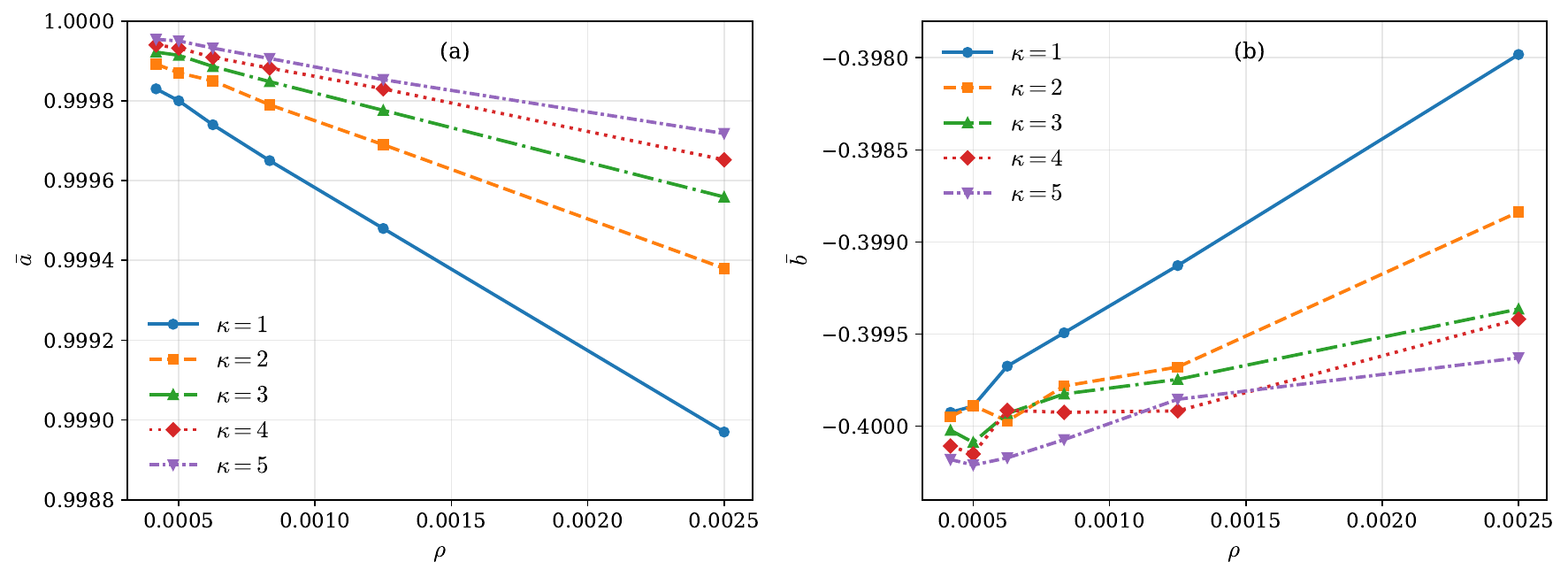}}
	\caption{Dependence of the strong deflection limit coefficients $\bar{a}$ and $\bar{b}$ on the Lorentzian--Euclidean black hole parameters $\rho$ and $k$.}
    \label{fig4}
\end{figure*}
Fig.~\ref{fig4} illustrates the dependence of the strong deflection limit coefficients $\bar{a}$ and $\bar{b}$ on the Lorentzian--Euclidean parameter $\rho$ for different values of the parameter $k$. Panel~(a) shows that the coefficient $\bar{a}$ decreases monotonically with increasing $\rho$ for all considered values of $k$. For a fixed $\rho$, larger values of $k$ yield higher values of $\bar{a}$, indicating that the parameter $k$ partially offsets the suppressing effect of $\rho$. Panel~(b) demonstrates that $\bar{b}$ increases with $\rho$, becoming less negative as $\rho$ grows. The increase is more pronounced for smaller values of $k$, while larger values of $k$ shift $\bar{b}$ toward lower values. These results indicate that both parameters significantly influence the strong-deflection limit coefficients that govern photon trajectories in the vicinity of the black hole.
\begin{figure*}
		\includegraphics[width=\columnwidth]{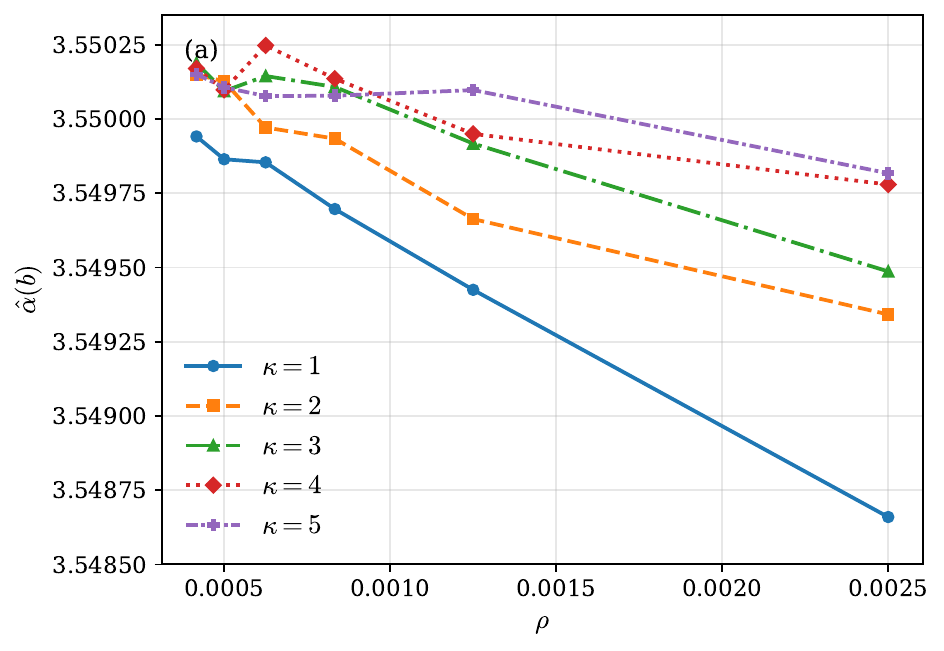}
	\caption{Variation of the strong-field deflection angle with respect to the Lorentz--Euclidean parameter $\rho$ for different values of $k$.}
    \label{fig5}
\end{figure*}
Fig.~\ref{fig5} presents the variation of the strong-field deflection angle $\alpha(b)$ as a function of the Lorentzian--Euclidean parameter $\rho$ for different values of $k$. The coefficient decreases monotonically with increasing $\rho$, indicating a gradual reduction in the strength of the logarithmic divergence characteristic of the strong deflection regime. For a fixed value of $\rho$, larger values of $k$ produce higher values of $a(b)$. The nearly linear dependence observed for all curves suggests that $\rho$ introduces a systematic correction to the strong-field lensing properties of the spacetime.
\begin{figure*}
	\centerline{
		\includegraphics[width=170mm,height=70mm]{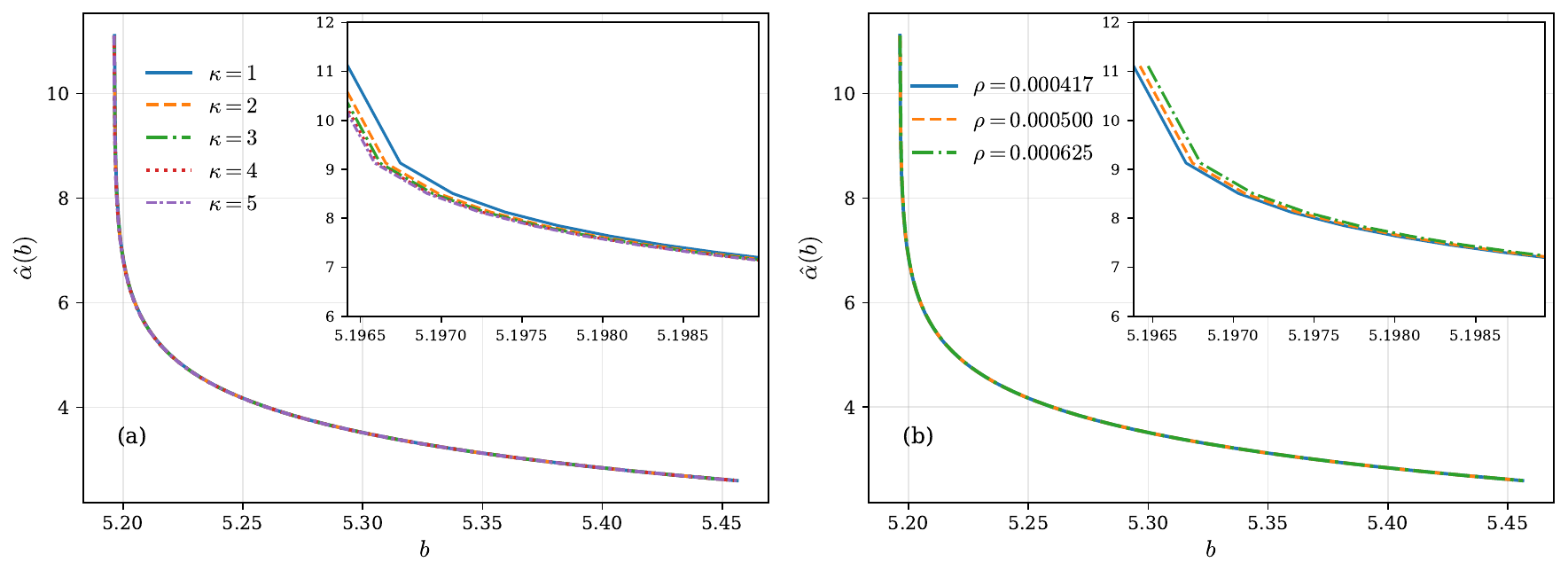}}
	\caption{The strong-field deflection angle as a function of the impact parameter $(b)$ for different values of $\rho$ and $k$.}
    \label{fig6}
\end{figure*}
Fig.~\ref{fig6} depicts the strong-field deflection angle $\alpha(b)$ as a function of the impact parameter $b$. Panel~(a) shows the influence of varying $k$ for a fixed value of $\rho$, whereas panel~(b) illustrates the effect of changing $\rho$ while keeping $k$ fixed. In both cases, the deflection angle decreases monotonically with increasing impact parameter and exhibits the characteristic divergence near the critical impact parameter associated with the photon sphere. The inset panels provide an enlarged view of the region close to the critical value, where the effects of $k$ and $\rho$ become more evident. Although the differences among the curves are small, both parameters induce measurable modifications to the lensing behavior in the strong-field regime.

\section{Lensing Observables in Strong deflection  Limit}\label{sec:5}
We aim to study strong gravitational lensing by evaluating lensing observables in the strong-field limit. To proceed, we present the lens equation in the strong-field limit. Lens geometry is illustrated in the following diagram
\begin{figure*}
\centering
{\includegraphics[width=8cm]{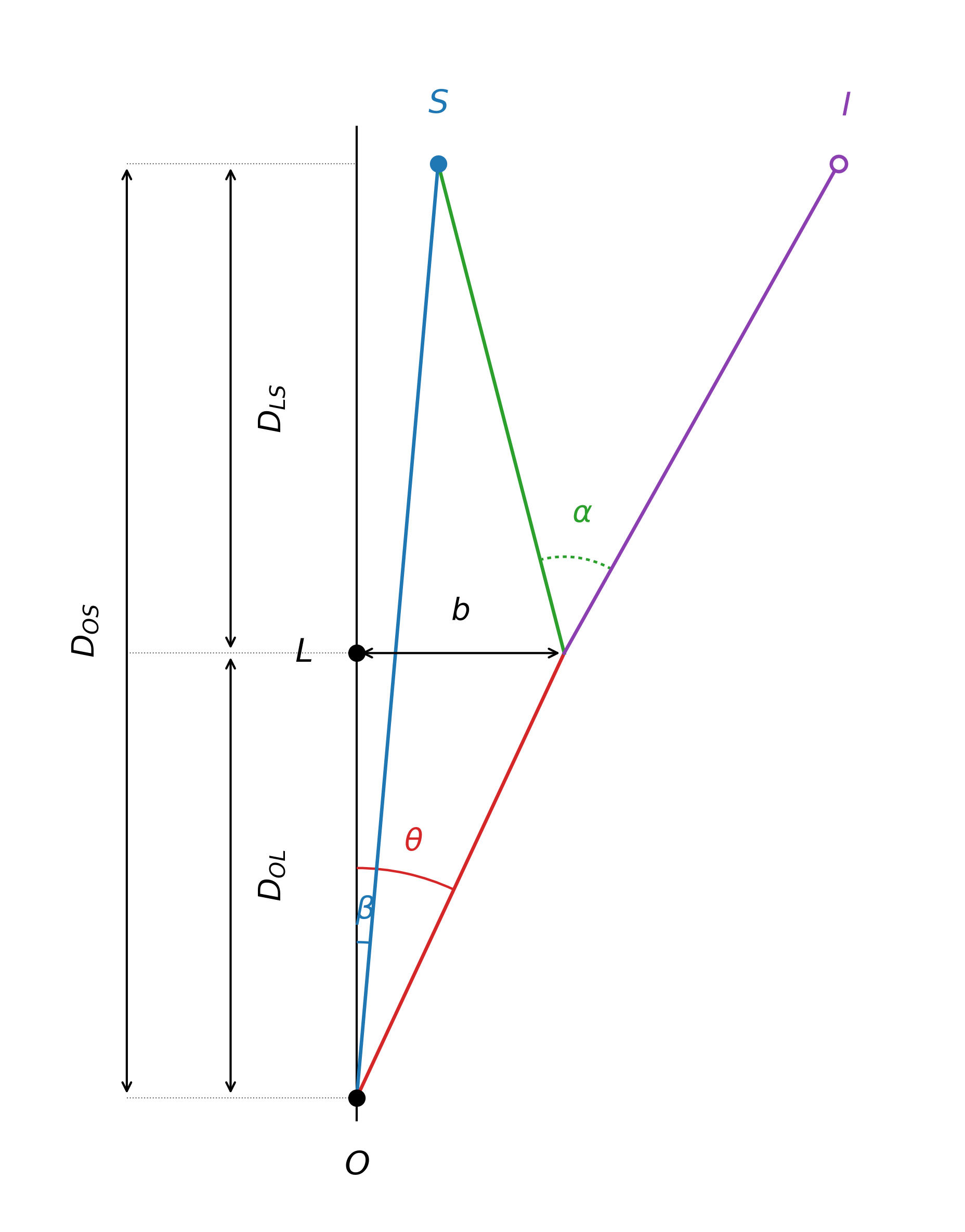}}
\vspace*{4pt}
\caption{Geometry of Lensing: A light
ray with  an impact parameter b emitted by a source S at a
source angle $\beta$ is deflected by a lens L with a deflection angle $\alpha$. The deflected ray is observed as an image I at an image angle $\theta$ by an observer O. $ D_{OS}$, $D_{LS}$, and $D_{OL}$ denote the distances between the observer and the source, between the lens and the source, and between the observer and the lens, respectively~\cite{tsukamoto2021gravitational}. \protect\label{fig7}}
\end{figure*}
Fig.~\ref{fig7} shows the schematic diagram of the lensing geometry. Light from the source S is deflected towards the observer O due to the presence of a gravitational lens. We define the standard quantities as follows:
$\beta$ denotes the angular position of the source, $\theta$ represents the angular position of the image, $\alpha$ denotes the bending angle, $D_{OL}$ is the distance between the observer and the lens, $D_{OS}$ is the distance between the observer and the source, and $D_{LS}$ is the distance between the lens and the source. The lens equation is then given by \cite{bozza2008comparison}
\begin{eqnarray}\label{Eq:47}
    \beta=\theta-\frac{D_{LS}}{D_{OS}}\Delta\alpha_{n},
\end{eqnarray}
where
\begin{eqnarray}
    \Delta\alpha_{n}=\alpha(\theta)-2n\pi,
\end{eqnarray}
represents the offset of the deflection angle after subtracting the contribution from all the loops completed by the photon \cite{bozza2002gravitational}. The deflection angle $\alpha(\theta)$ can be expressed as \cite{bozza2002gravitational}
\begin{equation}\label{Eq:49}
    \alpha(\theta) = -\overline{a}\log{\left(\frac{\theta D_{0L}}{b_{c}}-1\right)}+\overline{b}.
\end{equation}
To evaluate the offset angle $\Delta\alpha_{n}$ from $\alpha(\theta)$, we first determine the value of $\theta_{n}^0$, which corresponds to the angular position at which a light ray completes exactly $2n\pi$. This is found by imposing the condition $\alpha(\theta_{n}^0)=2n\pi$ in Eq.~(\ref{Eq:49}), which leads to
\begin{eqnarray}
  \theta_{n}^0=\frac{b_{c}}{D_{OL}}(1+e_{n}),
\end{eqnarray}
where
\begin{eqnarray}
e_{n}=\exp{\frac{(\overline{b}-2n\pi)}{\overline{a}}}.
\end{eqnarray}
Then, $\Delta\alpha_{n}$ can be obtained by expanding $\alpha(\theta)$ about $\theta=\theta_{n}^0$, with $\Delta\theta_{n}=\theta-\theta_{n}^0$, yielding
\begin{eqnarray}\label{Eq: 52}
    \Delta\alpha_{n} = -\frac{\overline{a}D_{OL}}{b_{c}e_{n}}\Delta\theta_{n}.
\end{eqnarray}
By substituting Eq.~(\ref{Eq: 52}) and $\theta=\theta_{n}^0+\Delta\theta_{n}$ into Eq.~(\ref{Eq:47}), we obtain
\begin{eqnarray}
 \beta=\theta^0_{n}+\Delta\theta_{n}+\left(\frac{\overline{a}D_{OL}}{b_{c}e_{n}}\frac{D_{LS}}{D_{OS}}\right)\Delta\theta_{n}.
\end{eqnarray}
Since
$b_{c}\ll D_{OL}$,
we neglect the second term. Hence, the position of the $n^{\rm th}$ image is given by \cite{bozza2002gravitational}
\begin{eqnarray}
\theta_{n}=\theta^0_{n}+\frac{b_{c}e_{n}(\beta-\theta^0_{n})D_{OS}}{\overline{a}D_{LS}D_{OL}}.\label{Eq:24}
\end{eqnarray}
We now define the magnification of the image as
\begin{equation}
\mu_n=
\left(
\frac{\beta}{\theta}
\frac{d\beta}{d\theta}
\Big|_{\theta=\theta_n^{0}}
\right)^{-1}.
\end{equation}
Its expression is given by
\begin{eqnarray}
    \mu_{n}=e_{n}\frac{b^2_{c}(1+e_{n})D_{OS}}{\overline{a}\beta D^{2}_{OL}D_{LS}}.
\end{eqnarray}
This relation indicates that the magnification decreases as $n$ increases. Consider a situation in which the outermost image, $\theta_{1}$, is resolved as a distinct image, while the remaining images are clustered together at $\theta_{\infty}$. Based on this, we define three lensing observables \cite{bozza2002gravitational}
\begin{eqnarray}
    \theta_{\infty} =\frac{b_{c}}{D_{OL}},
\end{eqnarray}
Then,
\begin{eqnarray}
   \overline{s}=\theta_{1}-\theta_{\infty} =\theta_{\infty}\exp{\left(\frac{\overline{b}-2\pi}{\overline{a}}\right)},
\end{eqnarray}
and
\begin{eqnarray}
   \mathcal{R}=\frac{\mu_{1}}{\sum_{N=2}^{\infty}{\mu_N}}
=\exp{\frac{2\pi}{\overline{a}}}.
\end{eqnarray}
Here,
$\overline{s}$ represents the angular separation between the outermost image and the remaining images, while
$\mathcal{R}$ is the flux ratio between the outermost image and the remaining images.

We now evaluate the time delay between the relativistic images, defined as the difference between the emission time of the source radiation and the reception time of the corresponding signal by the observer. In this work, we adopt the method proposed by Bozza and Mancini \cite{bozza2004time}. The time taken by a photon to travel from the source to the observer at infinity is given by \cite{bozza2004time}
\begin{eqnarray}
\overline{T}(b) =\overline{a}\log{\left(\frac{b}{b_{c}}-1\right)}+b_{c}+O(b-b_{c}).
\end{eqnarray}
Using this expression, we evaluate the time delay between the first and second relativistic images, both of which are formed on the same side of the source. The corresponding time delay is given by \cite{bozza2004time}
\begin{eqnarray}
\Delta T_{21} =2\pi b_{c} =2\pi D_{OL}\theta_{\infty}.
\end{eqnarray}
For the numerical estimation of lensing observables and time delays, we consider the supermassive black holes Sgr A* and M87* within the framework of the Lorentzian--Euclidean black hole spacetime. The mass of M87* is taken to be $(6.5 \pm 0.7)\times10^{9}M_{\odot}$, with a distance of $d=16.8\,\mathrm{Mpc}$~\cite{EventHorizonTelescope:2019ggy}. For Sgr A*, we adopt a mass of $(4.28 \pm 0.21 \pm 0.10)\times10^{6}M_{\odot}$ and a distance of $d=(8.32 \pm 0.07 \pm 0.14)\,\mathrm{kpc}$~\cite{Gillessen:2017jxc}. Following Moffat and Toth~\cite{Moffat:2019uxp}, the dynamical mass inferred from stellar orbital measurements is identified with the corresponding Newtonian mass that would be employed in Newtonian gravity and General Relativity. The numerical values of the strong gravitational lensing observables for SgrA$^{*}$ and M87$^{*}$ are presented in Table~\ref{tab:lensing} for different values of the model parameters $k$ and $\rho$. 

\begin{figure*}
	\centerline{
		\includegraphics[width=170mm,height=70mm]{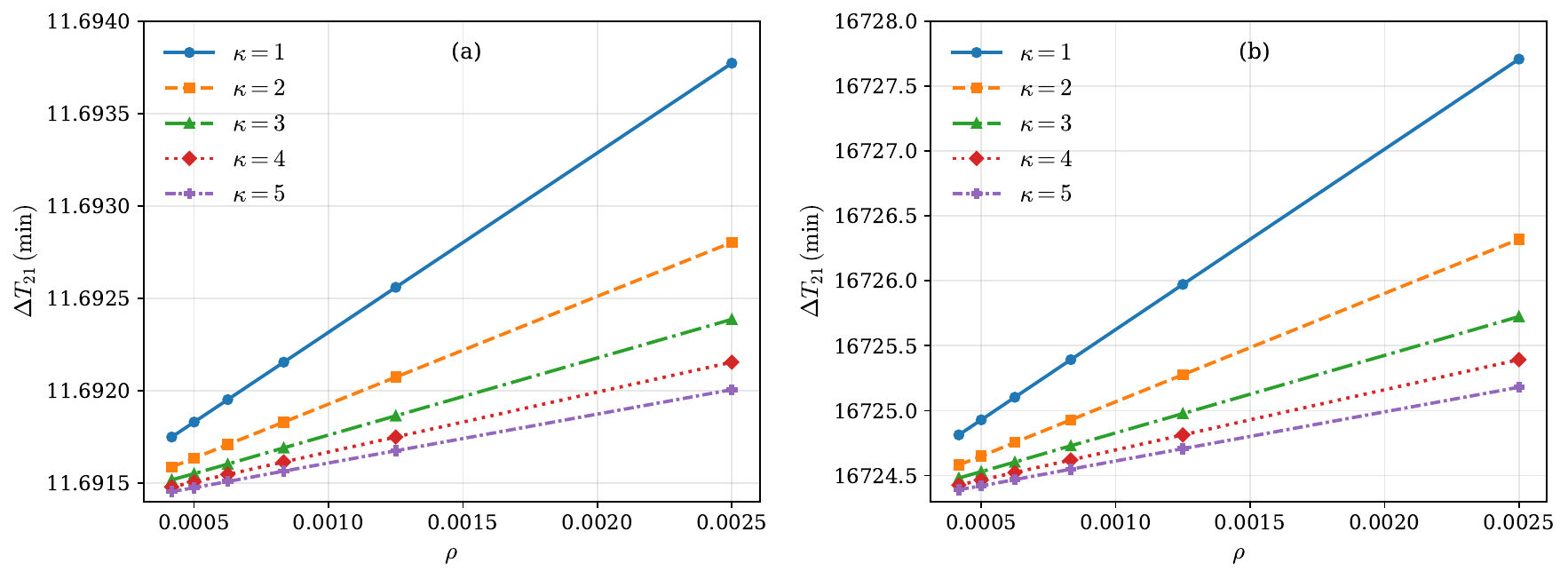}}
	\caption{Variation of the differential time-delay observable $\Delta T_{2,1}$ as a function of the Lorentzian--Euclidean parameter $\rho$ for different values of $k$. Panel (a) corresponds to Sgr A$^{*}$, while panel (b) corresponds to M87$^{*}$.}
    \label{fig8}
\end{figure*}

\begin{figure*}
	\centerline{
		\includegraphics[width=170mm,height=70mm]{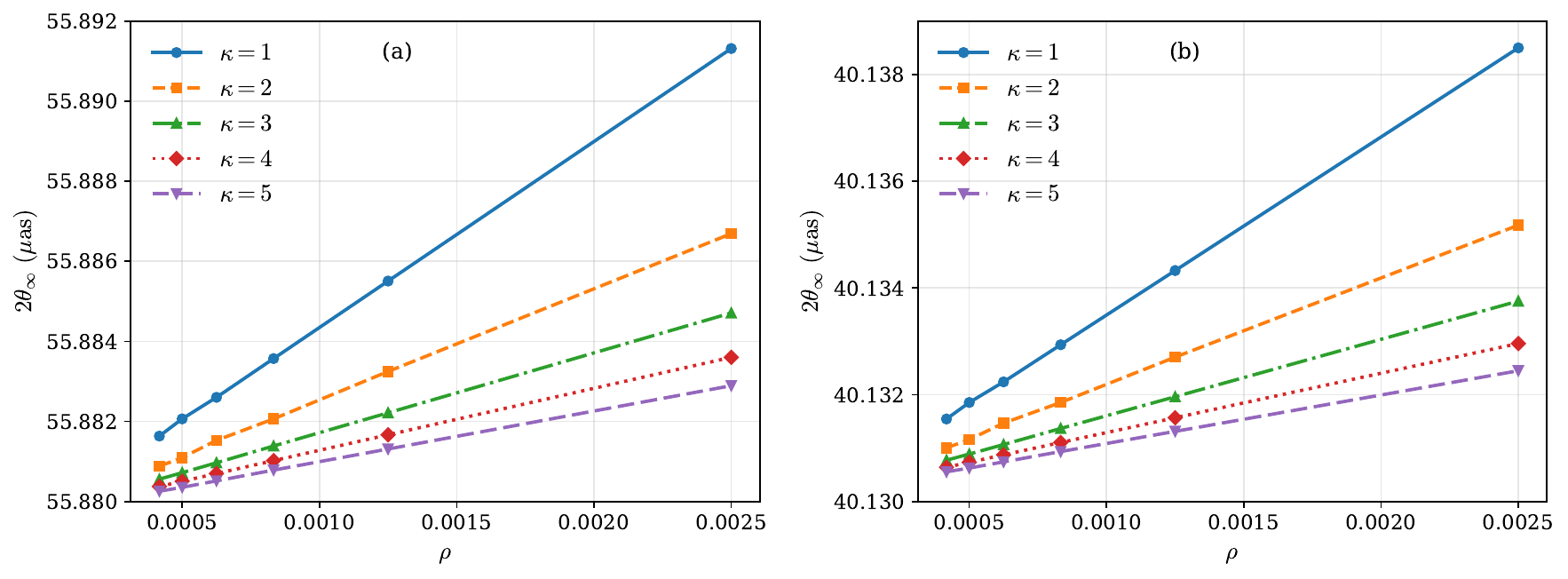}}
	\caption{Variation of the angular image position observable $2\theta_{\infty}$ as a function of the Lorentzian--Euclidean parameter $\rho$ for different values of $k$. Panel (a) corresponds to Sgr A$^{*}$, while panel (b) corresponds to M87$^{*}$.}
    \label{fig9}
\end{figure*}

\begin{figure*}
	\centerline{
		\includegraphics[width=170mm,height=70mm]{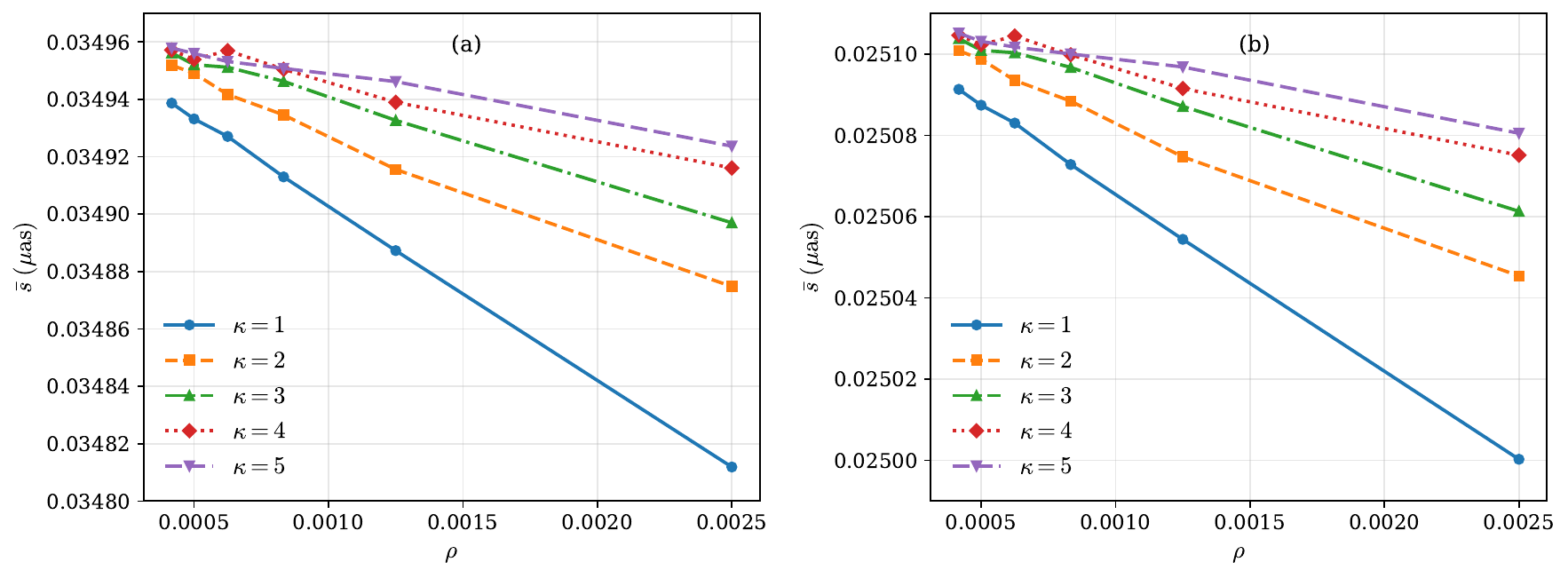}}
	\caption{Variation of the angular separation $\bar{s}$ $(\mu as)$ as a function of the Lorentzian--Euclidean parameter $\rho$ for different values of $k$. Panel (a) corresponds to Sgr A$^{*}$, while panel (b) corresponds to M87$^{*}$.}
    \label{fig10}
\end{figure*}

\begin{table*}
\caption{\label{tab:lensing}
Numerical estimation of lensing observables for SgrA* and M87 for different values of $k$ and $\rho$.}
\centering
\footnotesize
\setlength{\tabcolsep}{8pt}
\renewcommand{\arraystretch}{2.15}

\begin{tabular}{ccccccccc}
\hline\hline
$k$ & $\rho$ &
\multicolumn{3}{c}{SgrA*} &
\multicolumn{3}{c}{M87} &
$\mathcal{R}$ \\
\cline{3-5} \cline{6-8}

&
&
$2\theta_\infty$ $(\mu as)$ &
$\bar{s}$ $(\mu as)$ &
$\Delta T_{21}$ (min) &
$2\theta_\infty$ $(\mu as)$ &
$\bar{s}$ $(\mu as)$ &
$\Delta T_{21}$ (min) &
\\

\hline

\multirow{6}{*}{1}
& 0.000417 & 55.8816356& 0.0349387& 11.691749& 40.1315469 & 0.0250913 & 16724.81 & 536.06 \\
& 0.000500 & 55.8820657 & 0.0349332& 11.691830 & 40.1318558 & 0.0250874 & 16724.92 & 536.16 \\
& 0.000625 & 55.8826034 &0.0349271 & 11.691951 &40.1322420 & 0.0250830 & 16725.10 & 536.36 \\
& 0.000833 & 55.8835713 &0.0349130 & 11.692154& 40.1329371 & 0.0250728 & 16725.39 & 536.67 \\
& 0.001250 & 55.8855071 &0.0348873 & 11.692559 & 40.1343272 & 0.0250544 & 16725.97 & 537.24 \\
& 0.002500 & 55.8913142& 0.0348119 & 11.693771 & 40.1384977 & 0.0250003 & 16727.70 & 538.97 \\
\hline
\multirow{6}{*}{2}
& 0.000417 & 55.8808828 & 0.0349519& 11.691587& 40.1310061& 0.0251008
 & 16724.52 & 535.79 \\
& 0.000500 & 55.8810971& 0.0349491 & 11.691634 & 40.1311612 & 0.0250987 & 16724.60 & 535.89 \\
& 0.000625 & 55.8815282 &0.0349417 & 11.691708 & 40.1314697 & 0.0250935 & 16724.73 & 536.09 \\
& 0.000833 &55.8820651 &0.0349346 & 11.691830& 40.1318558 & 0.0250884 & 16724.95 & 536.40 \\
& 0.001250 & 55.8832487 &0.0349156& 11.692073 & 40.1327054 &0.0250747 & 16725.39 & 536.97 \\
& 0.002500 & 55.8866912 & 0.0348748& 11.692802 & 40.1351767 & 0.0250454 & 16726.69 & 538.70 \\
\hline
\multirow{6}{*}{3}
& 0.000417 &55.8805601& 0.0349562& 11.691517 & 40.1307746 &0.0251038
 & 16724.48& 535.75\\
& 0.000500 & 55.8807212 &0.0349521 & 11.691551 & 40.1308904 & 0.0251009 & 16724.53 &535.78 \\
& 0.000625 & 55.8809688 & 0.0349512 & 11.691603 & 40.1310681 &0.0251003 & 16724.60 & 535.87\\
& 0.000833 & 55.8813882 & 0.0349463& 11.691690& 40.1313693&0.0250967 &16724.72 & 536.01 \\
& 0.001250 & 55.8822163 & 0.0349327 & 11.691864 & 40.1319639 &0.0250871 & 16724.97 & 536.24 \\
& 0.002500 & 55.8847112 & 0.0348970 &11.692386 & 40.1337557 & 0.0250613 & 16725.72 &536.98 \\
\hline
\multirow{6}{*}{4}
& 0.000417 &55.8803773 & 0.0349572& 11.691479& 40.1306433 & 0.0251046
 & 16724.42& 535.69 \\
& 0.000500 & 55.8805064& 0.0349539 & 11.691506 & 40.1307360&0.02510230 & 16724.46 & 535.72 \\
& 0.000625 & 55.8806999& 0.0349570 & 11.691546 & 40.1308750 & 0.0251044& 16724.52 & 535.79\\
& 0.000833 &55.8810226& 0.0349505 & 11.691614 &40.1311067 & 0.0250998&16724.62 & 535.88 \\
& 0.001250 &55.8816678 &0.0349390 & 11.691749 & 40.1315701 & 0.0250915& 16724.81& 536.06\\
& 0.002500 &55.8836031& 0.0349161 & 11.692154 &40.1329602 & 
0.0250751 & 16725.39 & 536.66 \\
\hline
\multirow{6}{*}{5}
& 0.000417 & 55.8802590& 0.0349579 & 11.691454 & 40.1305583& 
0.0251051 & 16724.39&535.64\\
& 0.000500 & 55.8803558 & 0.0349560& 11.691474 & 40.1306278 & 0.0251031& 16724.42& 535.65 \\
& 0.000625 & 55.8805171&0.0349532 & 11.691508 & 40.1307437 & 0.0251017& 16724.46 &535.72 \\
& 0.000833 & 55.8807860 & 0.0349508 & 11.691564& 40.1309368 & 0.0251000 & 16724.54 & 535.80\\
& 0.001250 & 55.8813129& 0.0349462 & 11.691675 &40.1313152 &0.0250968 & 16724.71 &535.98 \\
& 0.002500 & 55.8828938 & 0.0349237 & 11.692005& 40.1324505 & 0.0250805 & 16725.18 & 535.98 \\
\hline\hline
\end{tabular}
\end{table*}

Fig.~\ref{fig8} shows the variation of the differential time delay $\Delta T_{2,1}$ with the Lorentzian--Euclidean parameter $\rho$ for different values of the parameter $k$. For both Sgr A$^{*}$ [panel (a)] and M87$^{*}$ [panel (b)], the differential time delay increases monotonically with increasing $\rho$. This behavior indicates that larger values of $\rho$ enhance the optical path length of photons executing multiple loops around the photon sphere before reaching the observer. For a fixed value of $\rho$, the time delay decreases as $k$ increases, suggesting that the parameter $k$ weakens the effective gravitational confinement of photon trajectories. Although the qualitative trends remain identical for both black holes, the absolute values of $\Delta T_{2,1}$ are substantially larger for M87$^{*}$ because of its significantly greater mass. The results imply that both $\rho$ and $k$ leave observable signatures on the temporal characteristics of strong gravitational lensing.\\

Fig.~\ref{fig9} depicts the dependence of the angular image position $2\theta_{\infty}$ on the parameter $\rho$ for different values of $k$. It is evident that $2\theta_{\infty}$ increases steadily with increasing $\rho$ for both Sgr A$^{*}$ and M87$^{*}$. This trend indicates an enlargement of the apparent photon ring diameter and reflects an increase in the critical impact parameter associated with the photon sphere. On the other hand, for a fixed value of $\rho$, increasing $k$ reduces $2\theta_{\infty}$, leading to a smaller angular extent of the relativistic image structure. The nearly linear dependence of $2\theta_{\infty}$ on $\rho$ demonstrates that the Lorentzian--Euclidean parameter strongly influences the angular scale of strong lensing observables. Consequently, measurements of the photon ring size can provide useful constraints on the parameters of the spacetime geometry.\\

Fig.~\ref{fig10} depicts the variation of the angular separation $s$ ($\mu\mathrm{as}$), one of the primary strong gravitational lensing observables, as a function of the Lorentzian--Euclidean parameter $\rho$ for different values of the parameter $k$. Panel~(a) corresponds to Sgr A*, while panel~(b) represents M87*. In both cases, the angular separation decreases monotonically with increasing $\rho$, indicating that the first relativistic image moves closer to the highly packed set of remaining relativistic images as the Lorentzian--Euclidean parameter increases. This behavior suggests that the relativistic images become progressively more compact, thereby reducing their observational resolvability.

The effect of $\rho$ is found to depend strongly on the parameter $k$. For smaller values of $k$, particularly $k=1$, the angular separation exhibits a steeper decline with increasing $\rho$, whereas larger values of $k$ significantly suppress this dependence, resulting in nearly convergent curves for $k=4$ and $k=5$. This trend indicates that increasing $k$ diminishes the influence of the Lorentzian--Euclidean parameter on the image configuration in the strong-field regime. Although both Sgr A* and M87* display identical qualitative behavior, the angular separation is consistently larger for M87* owing to its substantially larger gravitational radius. Since the observable $\bar{s}$ characterizes the angular distance between the outermost relativistic image and the remaining unresolved image cluster, its systematic decrease with $\rho$ implies that resolving the relativistic images becomes increasingly challenging for larger values of the Lorentzian--Euclidean parameter. Therefore, precise measurements of $s$ from future very-long-baseline interferometric observations could provide valuable constraints on both the Lorentzian--Euclidean parameter $\rho$ and the spacetime parameter $k$, thereby offering a potential observational test of the underlying black hole geometry.

\section{Constraints from EHT Observations of Sgr A* and M87*}\label{sec7}

To assess the astrophysical viability of the Lorentzian--Euclidean black hole as a candidate for real supermassive compact objects, we confront the model with the shadow observations of Sgr A* and M87* reported by the Event Horizon Telescope (EHT) collaboration. The angular shadow diameter is expressed through the null photon-sphere structure of the metric via the standard relation $2\theta_\infty = 2\,b_{ph}\,(GM/c^2)/D$, where $b_{ph}$ is the critical impact parameter associated with the unstable circular photon orbit, and $M$, $D$ are the mass and distance of the source. We use $M=(6.5\pm0.7)\times10^{9}M_\odot$, $D=16.8\,$Mpc for M87*, and $M=(4.28\pm0.21\pm0.10)\times10^{6}M_\odot$, $D=(8.32\pm0.07\pm0.14)\,$kpc for Sgr A*, consistent with the values adopted in Sec.~\ref{sec:2} for the lensing analysis. The EHT-inferred shadow angular diameters, $\theta_{sh}=48.7\pm7\,\mu$as for Sgr A* and $\theta_{sh}=42\pm3\,\mu$as for M87*, define the $1\sigma$ observational band against which the model prediction $2\theta_\infty(\rho,k)$ is tested.

\begin{figure*}
	\centerline{
		\includegraphics[width=170mm,height=80mm]{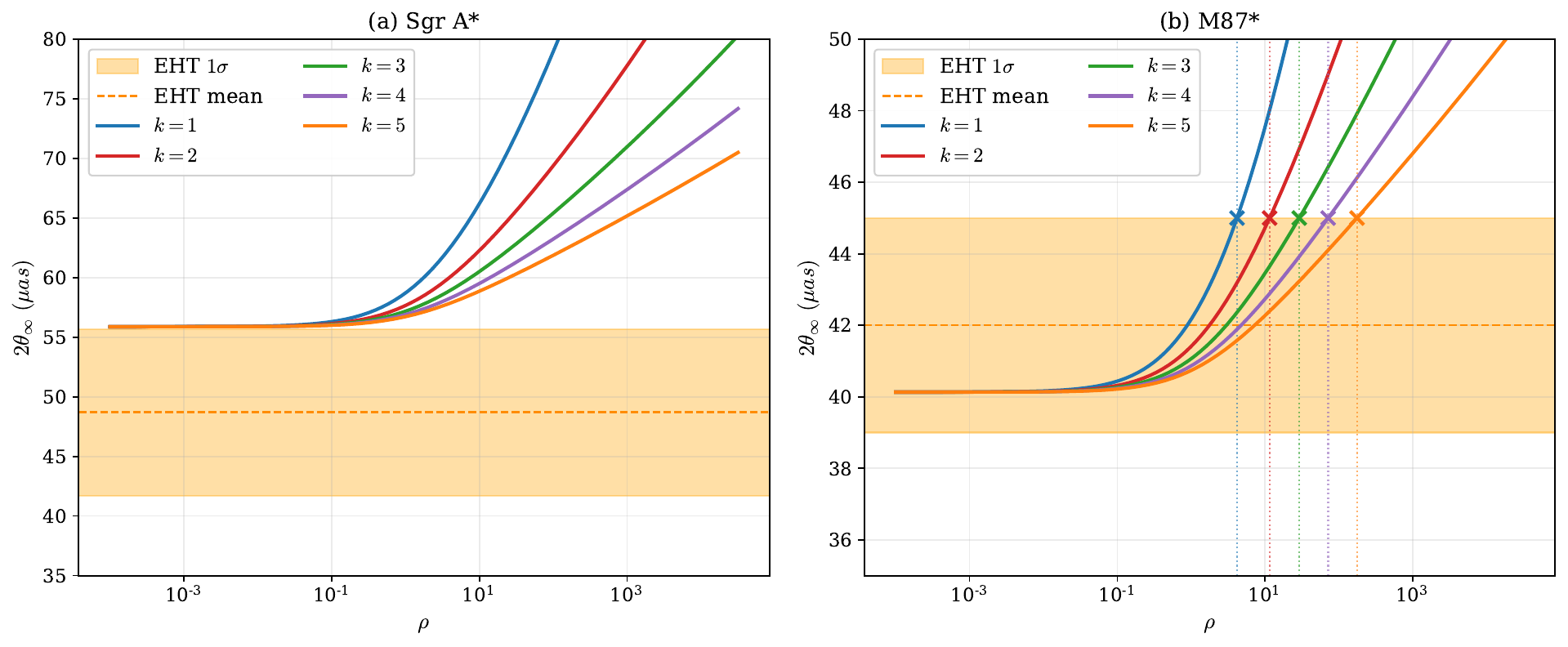}}
	\caption{Variation of the angular image-position observable $2\theta_\infty$ (in $\mu$as) with the Lorentzian--Euclidean regularization parameter $\rho$ (in units of $M^2$), shown on a logarithmic horizontal scale, for diffent values of $k$. Panel (a) corresponds to Sgr~A* and panel (b) to M87*. The shaded orange band marks the EHT $1\sigma$ observational interval for the shadow angular diameter, $\theta_{sh}=48.7\pm7\,\mu$as for Sgr~A* and $\theta_{sh}=42\pm3\,\mu$as for M87*, with the dashed orange line indicating the central value. }
    \label{fig11}
\end{figure*}

Fig.~\ref{fig11} shows the variation of $2\theta_\infty$ with $\rho$ for $k=1,2,3,4,5$, extended well beyond the fiducial parameter range used in Table~\ref{tab:lensing}, together with the EHT $1\sigma$ shaded band. Panel (a) corresponds to Sgr A* and panel (b) to M87*. In both cases $2\theta_\infty$ increases monotonically with $\rho$ for fixed $k$, and decreases with increasing $k$ for fixed $\rho$, reflecting the fact that a larger regularization width and/or a slower ($k\to1$) approach of $\epsilon(r)$ to $\pm1$ allows the near-horizon modification to reach further into the photon-sphere region at $r\simeq3M$ and thereby perturb the shadow more strongly.

For Sgr A*, the Schwarzschild limit ($\rho\to0$) already lies marginally above the $1\sigma$ upper edge of the observed band ($2\theta_\infty\simeq55.88\,\mu$as versus the band $[41.7,55.7]\,\mu$as) for every value of $k$ considered, and increasing $\rho$ only pushes the prediction further from the band. Consequently the model does not fall within the Sgr A* $1\sigma$ interval for any $\rho\geq0$; this source therefore does not, at present precision, provide a two-sided constraint on $\rho$, but instead indicates a mild, $k$-independent tension at the $1\sigma$ level that is common to the entire regularization family, including its Schwarzschild limit.

For M87*, by contrast, the Schwarzschild limit already lies inside the $1\sigma$ band ($2\theta_\infty\simeq40.13\,\mu\mathrm{as}$ versus the band $[39,45]\,\mu\mathrm{as}$). Consequently, the model is consistent with the $1\sigma$ EHT constraint at $\rho=0$ and remains so up to a finite upper bound on $\rho$, whose values for different choices of $k$ are summarized in Table~\ref{tab:m87_constraint}. Beyond these limits, the predicted shadow diameter exceeds the upper edge of the $1\sigma$ observational interval for M87*. The near-geometric growth of the allowed upper bound with increasing $k$ (each unit increase in $k$ enlarges the permissible range of $\rho$ by a factor of $\sim2.5$--$3.5$) reflects the increasingly efficient suppression of the near-horizon modification by $\epsilon(r)$, thereby reducing its influence on the photon-sphere region.

\begin{table}[h]
\centering
\caption{Upper bounds on the Lorentzian--Euclidean parameter $\rho$ for M87* obtained from the $1\sigma$ EHT constraint on the shadow diameter.}
\label{tab:m87_constraint}
\begin{tabular}{cc}
\hline
$k$ & Maximum allowed $\rho$ \\
\hline
1 & $\rho \lesssim 4.17$ \\
2 & $\rho \lesssim 11.51$ \\
3 & $\rho \lesssim 28.97$ \\
4 & $\rho \lesssim 71.38$ \\
5 & $\rho \lesssim 175.3$ \\
\hline
\end{tabular}
\end{table}

The fiducial values $\rho=0.000417$--$0.0025$ adopted in Table~\ref{tab:lensing} lie three to five orders of magnitude below all of the M87* derived bounds quoted above, confirming that the strong-lensing observables reported there are computed in a regime that is comfortably $1\sigma$-compatible with M87*, and subject only to the mild $\rho$-independent Sgr A* tension noted above. The results of this section are therefore complementary rather than restrictive for the fiducial lensing study: they establish the outer edge of parameter space still allowed by present M87* shadow data, well beyond which the working values of $(\rho,k)$ are safely nested.

\subsection{Observational Implications and Parameter Degeneracy}

These bounds provide a genuine and falsifiable observational test of the horizon regularization scenario. The M87* shadow size responds monotonically and appreciably to $\rho$ once the transition width becomes comparable to the horizon to photon sphere separation ($\sim M$). Consequently, the current EHT data already exclude regularization widths of order ten and larger for $k=1$. The excluded region decreases rapidly as $k$ increases. This is a nontrivial result because the regularization was introduced solely to remove the formal curvature singularities at the signature changing surface $r=2M$ \cite{Capozziello:2024ucm,DeBianchi:2025bgn,Capozziello:2025wwl}. The shadow analysis demonstrates that this regularization remains consistent with the M87* observations at the $1\sigma$ confidence level, provided $\rho\lesssim\mathcal{O}(1\text{ to }10^2)$ (in units of $M^2$), depending on $k$. The persistent and $k$ independent mild tension with Sgr A* is already present in the Schwarzschild limit of the model. This serves as a useful diagnostic because it indicates that the discrepancy originates from the underlying Schwarzschild like optical geometry common to the entire family rather than from the regularization parameters themselves. Therefore, no choice of $(\rho,k)$ removes this tension. As next generation EHT and ngVLA observations reduce the uncertainty in $\theta_{sh}$ for both sources, the bounds derived here will become increasingly stringent, thereby providing a concrete observational probe of the parameters governing the horizon regularization.

We emphasize that the shadow observable alone constrains only a combination of $\rho$ and $k$, not the two parameters independently. As is evident from Fig.~\ref{fig11}, an increase in $\rho$ (wider transition region) can be compensated by an increase in $k$ (sharper healing profile) while leaving $2\theta_\infty$ essentially unchanged; the two parameters enter the regularized function $\epsilon(r)$ only through the combination that controls how far the deviation from $\mathrm{sign}(r-2M)$ extends into the region $r\gtrsim2M$, and the shadow is sensitive only to the net effect of this combination at $r\simeq3M$. Consequently, the exclusion boundary reported above for M87* should be interpreted as a degenerate contour in the $(\rho,k)$ plane rather than as an independent bound on each parameter: any pair $(\rho,k)$ lying on (or inside) this contour is statistically indistinguishable from any other pair on the same contour using the shadow diameter alone. Breaking this degeneracy requires additional observables beyond the angular size of black hole shadow. The shadow shape, relativistic image time delays, and quasinormal-mode spectra may provide complementary constraints on $\rho$ and $k$. We leave such a combined analysis for future work.

\section{Conclusion}\label{sec:6}
In this work, we have presented a systematic study of strong gravitational lensing and shadow observables for the Lorentzian--Euclidean black hole, a spacetime in which the Schwarzschild horizon is replaced by a regular signature changing surface characterized by the parameters $\rho$ and $k$. Although this geometry has been studied mathematically, and its black hole shadow was recently investigated by Battista \emph{et al.}~\cite{battista2026shadow}, its gravitational lensing signatures and observational constraints on the model parameters remain unexplored.

The null geodesic analysis shows that the photon sphere radius and critical impact parameter increase with $\rho$ and decrease with $k$, demonstrating that the two parameters produce opposite effects on the optical geometry. Using the strong deflection limit, we calculated the deflection angle, strong-lensing coefficients, angular position of the relativistic images $\theta_\infty$, angular separation $\bar{s}$, relative magnification $r_{\rm mag}$, and differential time delay $\Delta T_{2,1}$ for Sgr A* and M87*. The results show that $\theta_\infty$ and $\Delta T_{2,1}$ increase with $\rho$ and decrease with $k$, leading to observable deviations from the Schwarzschild case that may be tested with future VLBI observations. In contrast, the angular separation $\bar{s}$ decreases with increasing $\rho$, with its dependence becoming progressively weaker for larger values of $k$. Consequently, precise measurements of $\bar{s}$ could provide valuable observational constraints on the Lorentzian--Euclidean parameters $\rho$ and $k$.

We compare the model predictions with the $1\sigma$ EHT shadow observations to derive observational constraints on the model parameters. For Sgr A*, the Schwarzschild limit already lies slightly outside the observed $1\sigma$ interval for all values of $k$, whereas for M87* it remains within the allowed range and yields a $k$ dependent upper bound on the regularization parameter $\rho$. The allowed value of $\rho$ increases from approximately $4.2$ for $k=1$ to about $175$ for $k=5$. The parameter values adopted in our numerical analysis lie well below these limits, confirming that the predicted strong lensing observables correspond to an astrophysically viable region of the parameter space. 

Furthermore, the combined effects of the parameters $\rho$ and $k$ were investigated. Since $\rho$ determines the width of the transition region and $k$ controls its sharpness, different combinations of these parameters can produce nearly identical shadow diameters. The present constraints should therefore be regarded as limits on their combined effect. Additional observables, such as higher order relativistic images, and quasinormal mode spectra, are expected to break this degeneracy and provide independent constraints on the two parameters.

The present analysis shows that the Lorentzian--Euclidean regularization is consistent with the current $1\sigma$ EHT observations of M87* over a broad parameter range while exhibiting a mild tension with Sgr A*. Future observations with next generation EHT facilities, space VLBI, and gravitational wave detectors will provide significantly improved measurements of black hole shadows and strong field observables, enabling more stringent tests of this regularized spacetime and its near horizon geometry.


\section*{Acknowledgments}
The authors would like to thank Emmanuele Battista for useful discussions. The authors RVP and SR gratefully acknowledge the support of DBT Star College Scheme and DST-FIST (SR/FST/College-139/2013 c) for providing research facilities at Providence Women’s College, Calicut, India. SK sincerely acknowledges IMSc for providing exceptional research facilities and a conducive environment that facilitated his work as an Institute Postdoctoral Fellow.

\bibliographystyle{unsrt}
\bibliography{mainLEBH}

\end{document}